\documentclass[conference]{IEEEtran}
\IEEEoverridecommandlockouts
\usepackage{cite}
\usepackage{amsmath,amssymb,amsfonts}
\usepackage{graphicx}
\usepackage{algorithmic}
\usepackage[]{algorithm2e}
\usepackage{textcomp}
\usepackage{xcolor}
\def\BibTeX{{\rm B\kern-.05em{\sc i\kern-.025em b}\kern-.08em
    T\kern-.1667em\lower.7ex\hbox{E}\kern-.125emX}}
\begin{document}

\title{Explaining Differences in Classes\\ of Discrete Sequences
}

\author{\IEEEauthorblockN{Samaneh Saadat}
\IEEEauthorblockA{\textit{Department of Computer Science} \\
\textit{University of Central Florida}\\
Orlando, FL US \\
ssaadat@cs.ucf.edu}
\and
\IEEEauthorblockN{Gita Sukthankar}
\IEEEauthorblockA{\textit{Department of Computer Science} \\
\textit{University of Central Florida}\\
Orlando, FL US \\
gitars@eecs.ucf.edu}

}


\maketitle

\begin{abstract}
While there are many machine learning methods to classify and cluster sequences, they fail to explain what are the differences in groups of sequences that make them distinguishable. Although in some cases having a black box model is sufficient, there is a need for increased explainability in research areas focused on human behaviors. For example, psychologists are less interested in having a model that predicts human behavior with high accuracy and more concerned with identifying differences between actions that lead to divergent human behavior. This paper presents techniques for understanding differences between classes of discrete sequences. Approaches introduced in this paper can be utilized to interpret black box machine learning models on sequences. The first approach compares k-gram representations of sequences using the silhouette score. The second method characterizes differences by analyzing the distance matrix of subsequences. As a case study, we trained black box supervised learning methods to classify sequences of GitHub teams and then utilized our sequence analysis techniques to measure and characterize differences between event sequences of teams with bots and teams without bots. In our second case study, we classified Minecraft event sequences to infer their high-level actions and analyzed differences between low-level event sequences of actions. 
\end{abstract}


\section{Introduction}
Discrete event sequences are abundant in the online world: from user clicks in an online store to actions of users on social media platforms.
Researchers across different fields are interested in studying these digital traces, from social scientists who wish to improve their understanding of human cognition to computer scientists who are eager to develop methods that improve people's lives.
While there are machine learning approaches available for classifying and clustering sequences, there is a lack of approaches that determine ``what'' are the differences between groups of sequences.
This paper introduces a new analytic approach to characterizing differences between groups of event sequences. We aim to answer the following questions: (1) How distinct are the sequences of different groups? (2) What are the differences between these sets of sequences?

This paper demonstrates the usage of our analytic techniques on two different platforms: GitHub (the social coding site) and Minecraft (a massively multiplayer online game).
First, we review several approaches that can be used for classifying discrete sequences.  
We apply these methods to event sequences to classify whether a software engineering team uses GitHub automated services (bots) and to identify high-level Minecraft player behaviors. 
Second, we propose analytic approaches to understand the differences between groups of discrete sequences.
These approaches were used to compare team sequences in GitHub (bot vs.\ non-bot), as well as different actions in the Minecraft game.
 We modified the matrix profile algorithm~\cite{yeh2017matrix} to make it compatible with sequences. In matrix profiles, subsequences of the time series are compared against themselves by computing the distance between each pair of subsequences. The output of the matrix profile is then used to detect whether characteristics of groups of sequences such as complexity and novelty are different across groups.

This paper presents a case study comparing GitHub team event sequences of teams with and without automated services.
GitHub is a social coding platform that facilitates distributed, asynchronous collaborations in open source software (OSS) development.
GitHub provides developers with automated services (aka bots) through GitHub Marketplace, the platform’s store for development tools. 
It offers a comprehensive set of tools to support software development by virtual teams which makes it an ideal laboratory for studying teamwork. 
Code development, issue reporting, and social interactions are tracked by the 10+ event types.  Our assumption is that each software repository is maintained by a team and that the events associated with the repository form a partial history of the team activities and social interactions.

Our second case study analyzes Minecraft action sequences. Minecraft is a massively multiplayer online game, where players can explore a 3D world, mine materials, and craft tools and structures.
Game events form sequences that provide valuable information about the play style and high level goals of the players. 
The observable events are low-level: move, place block, consume item, etc. High-level actions in the game world, such as exploring, mining, fighting, or building, are accomplished by performing chains of low-level actions.  Since events are logged every few seconds, the sequence of low-level game events may be long and filled with superfluous detail.
Prior research attempted to classify these low-level event sequences to high-level actions \cite{muller2015statistical}.


\section{Related Work}
\subsubsection{Explainability}
Many of the most accurate machine learning models are constructed as black boxes, meaning that their internal logic is hidden from their users~\cite{guidotti2018survey}.
In the scientific community, there is an increasing interest in explaining decisions made by black box models.
For example, Guidotti et al. \cite{ECML2019Guidotti} presented an approach for explaining black box decisions of an image classification model.
Despite the abundance of discrete sequences, explaining machine learning models built for sequences has not been well-studied.

 \subsubsection{Discriminative Sequences} 
 One of the conventional approaches for finding key sequences is sequential pattern mining~\cite{fournier2017survey}, and researchers have proposed many specific techniques for discovering discriminative patterns that occur at significantly different frequencies across two groups of sequences~\cite{he2019significance,he2019mining,du2016discriminative}.  A related technique, motif mining, has been successfully applied to finding recurring sub-structures in call graphs~\cite{russo2018}.  A major drawback of sequential pattern mining approaches is the overabundance of sequences they generate as output.
 This makes large-scale sequence mining challenging for sequences for which we do not have prior knowledge.
Moreover, interpreting discovered patterns in sequential pattern mining requires subject-matter expertise. 
Our approach summarizes sequences and their differences into a few numbers, reducing the need for domain knowledge to interpret the data.  

\subsubsection{Matrix Profile}
Discrete sequences are the categorical analog of time series data.
Many time series analysis algorithms can be adopted to study discrete sequences.
In this paper, we utilize an algorithm to construct matrix profiles, introduced by Yeh et al. \cite{yeh2016matrix}, as a tool for conducting a large-scale analysis of sequences.  
The matrix profile is a vector calculated between two time series (similarity join) or one time series and itself (similarity self-join); for each subsequence of the first time series, the distance is stored to its closest subsequence in the second time series.
The distance between two subsequences is their Euclidean distance.
Matrix profiles have many applications in time series analysis including motif discovery, discord/anomaly detection, and semantic segmentation \cite{yeh2016matrix,yeh2018time,yeh2017matrix,torkamani2017survey}.
In this study, we created the matrix profile for discrete sequences and use it to summarize differences between groups of sequences. 

Yeh et al. \cite{yeh2016matrix} applied the matrix profile to DNA sequences but they convert the sequences to time series first using a method proposed by Rakthanmanon et al. \cite{rakthanmanon2012searching}.
This method converts sequences to time series by assigning an ordinal number to each symbol in the sequence.
Then, Euclidean distance is used for measuring the distance between time series.
With this ordinal encoding, an ordinal relationship is considered between symbols while in reality there may not be any relationship between symbols.
We developed an approach that extracts the matrix profile directly from discrete sequences. 
Instead of Euclidean distance, Hamming distance is employed; this distance measure is designed for sequences and does not require encoding of the input.

\subsubsection{GitHub}
The most commonly employed methodology for studying collaborative work in GitHub is an examination of the activity profiles of developers and/or their responses to surveys and semi-structured interviews to better understand important processes and outcomes of interest \cite{saadat2018initializing,blincoe2016understanding}.
Previous GitHub research has used features such as code comments~\cite{kavaler2017perceived} and issue closure~\cite{jarczyk2018surgical} to analyze team dynamics.  The aim of our research is to leverage event sequences generated by the team members to understand teamwork differences.
 
 There is very little research on the effect of these GitHub bots on software engineering teams ``in the wild.''  This paper introduces a set of analytic tools for quantifying teamwork differences between human-bot and human-only teams over a larger population than can be recruited for most lab studies.
 
\subsubsection{Minecraft}
Although Minecraft was not explicitly developed for research purposes, it has been used in many studies on learning and collaboration~\cite{nebel2016mining,debkowski2016contained}.
Muller et al.~\cite{muller2015statistical} studied players' actions using the frequencies of low-level events, in contrast to our work that leverages sequence information. 
Saadat et al.~\cite{saadat2020contrast} analyzed Minecraft event sequences and discovered contrast motifs of Minecraft actions and players.

\section{Method}
This section describes our techniques for analyzing event sequences.
First, the process of distinction measurement using silhouette score is described. Second, our approach for characterizing differences across groups of sequences is explained. Finally, we discuss the sequence classification methods we applied to our case study.
The code for our analytic pipeline is publicly available at \cite{supp}.

\subsection{Sequence Distinction Measurement Using Silhouette Score} \label{ss:kgram}
Given two groups of sequences, our goal in this section is to measure and characterize the distinctions. 
We model the problem of comparing groups of sequences as a clustering evaluation problem as they have similar purposes.
In a clustering evaluation problem, the goal is to measure the similarity of instances within and across clusters.
In a well-performed clustering, instances in a cluster are closely related (cluster cohesion) and instances of one cluster are distinct from other clusters (cluster separation).
Our goal is also to understand and measure cohesion and separation of groups of sequences.
Hence we used a clustering evaluation method, called silhouette score, to measure the distinction of two groups of sequences.
To calculate silhouette score, we need to have sequences in an equal length vector format. 
This section describes the sequence to vector conversion process and the silhouette score computation procedures.

\subsubsection{Sequence to Vector} \label{sss:seq2vec}
The first step in making sequences comparable is converting them to equal length vectors. 
For vectorization, we extract a k-gram representation of sequences by moving an overlapping window with fixed size of $w$ and step size of $1$ along the sequence to generate $n=l-w+1$ subsequences where $l$ is the length of the sequence.
This transformation converts each sequence to a set of ordered subsequences of length $w$.
To be able to compare a k-gram representation of sequences with each other, we convert them to vectors with equal lengths using the \textit{Term Frequency-Inverse Sequence Frequency} (\textit{TF-ISF}) model.
\textit{TF-ISF} is analogous to the \textit{TF-IDF} procedure that is used for vectorizing textual documents (i.e. an ordered list of words) \cite{manning2010introduction}.
Term frequency (\textit{TF}) measures the frequency of every subsequence in a sequence.
Higher frequency subsequences tend to contribute noise to the similarity computation \cite{aggarwal2015data}. 
One way to avoid this noise is by lowering the weighting subsequences with higher frequency using inverse sequence frequency (\textit{ISF}).
If there is a subsequence that is shared between most of the sequences, this subsequence may be less important in understanding differences between groups of sequences. 
The inverse sequence frequency $\mbox{ISF}_i$ of the \textit{i-th} subsequence is calculated using Equation \ref{eq:idf}.   
\begin{equation} \label{eq:idf}
    \mbox{ISF}_i=\log(N/N_i)
\end{equation}
where $N$ is the total number of sequences and $N_i$ is the number of sequences that contain the \textit{i-th} subsequence. Note that $\mbox{ISF}_i$ is a decreasing function of the number of sequences in which it occurs.

To summarize, in order to create the vector of sequences  using \textit{TF-ISF} model, we first measure the frequency of each subsequence $i$ ($\mbox{TF}_i$) and then multiply them by $\mbox{ISF}_i$. A subsequence has a high \textit{TF-ISF} for a sequence if it appears many times in that sequence and does not appear in many other sequences \cite{rajaraman2011mining}.

In addition to the \textit{TF-ISF} model for vectorization, we have a simple binary vectorization method.
In this model, each vector has zero and one values where one indicates the existence of a subsequence in the sequence and zero otherwise (i.e. frequencies of subsequences are ignored).
Comparing this model with the \textit{TF-ISF} model is helpful in understanding whether the difference between two groups of sequences is solely due to difference in the frequencies or whether the subsequences also differ.

\subsubsection{Sequence Comparison}
To compare two groups of sequences, we need a distance measure and also a metric that tells us the amount of similarity (or dissimilarity) between two groups.
We use cosine similarity and silhouette score for these purposes, respectively.

Silhouette score is mainly used to measure how well a set of samples is clustered and to compare the results of different clustering methods or configurations \cite{rousseeuw1987silhouettes}.
In this paper, silhouette score is used to measure the relative distinctiveness of two groups of sequence vectors. The silhouette score is calculated using the mean intra-group distance and the mean nearest-group distance for each sequence.

The best value of silhouette score is $1$ and the worst value is $-1$.
A score of $1$ indicates that two groups are completely separate. 
Values near zero show that groups are overlapping. 
Negative values generally indicate that sequences of one group will be incorrectly assigned to the other group.  If the silhouette score between the vectors of the two groups is higher, they are more distinct from each other. We calculate silhouette score to measure the separation between groups of event sequences.

Investigating variations in vectorization of sequences can inform us about the underlying differences in sequences.
In \textit{TF-ISF}, frequencies of k-grams are considered and in the binary version, only the existence of k-grams is investigated.
For example, if silhouette score detects no distinction based on binary vectors but denotes distinction based on the \textit{TF-ISF}, it means that top k-grams are similar across different groups but they have different frequencies.

\subsection{Sequence Difference Detection Using Matrix Profile} \label{ss:matrix_profile}
The vectorization method explained in Section \ref{sss:seq2vec} is good for large-scale comparison of sequences and reveals the differences in short and exact subsequeneces.
However, this vectorization method is computationally expensive for comparing large subsequences.
This section illustrates how we use matrix profiles to summarize differences between two groups of sequences based on large subsequences. 

The first step in the creation of the matrix profile is constructing the distance matrix $D$ which is a $n \times n$ matrix where $D[i, j]$ represents the distance between \textit{i-th} and \textit{j-th} subsequences.

In this paper, we used Hamming distance because of its computational efficiency. However, other distance calculation methods designed for sequences, such as Longest Common Subsequence (LCS), can be used.
Hamming distance is a simple distance function that calculates the number of mismatching positions between two sequences of equal length.
The Hamming distance calculation is fast with a time complexity of $O(w)$ where $w$ is the length of its input sequence. 
The distance matrix calculation time complexity is $O(n^2w)$ if Hamming distance is used.

For efficiency reasons, Yeh et al. \cite{yeh2016matrix} skip calculating  the 2-dimensional distance matrix. However in this study, our main purpose is providing tools that facilitate pattern discovery within and across sequences, hence it is helpful to store and visualize the 2-dimensional distance matrix.
Algorithm~\ref{alg:dist_mat} shows the distance matrix calculation procedure; $\mbox{distance}(i,j)$ is the distance between subsequence $i$ and subsequence $j$ which can be calculated using Hamming or LCS-based distance.

\begin{algorithm}
 \begin{algorithmic}[1]
 \renewcommand{\algorithmicrequire}{\textbf{Input:}}
 \renewcommand{\algorithmicensure}{\textbf{Output:}}
 \REQUIRE $s$: input sequence and $w$: window size
 \ENSURE  $D$: pairwise distance between subsequences
 \\ 
  \STATE $l = length(s)$
  \STATE $n = l - w + 1$
  \STATE  $D = n \times n$ matrix with default values of $w$
  \STATE $r = w / 2$
  \FOR {$i = 0$ to $n$}
    \FOR {$j = i + r$ to $n$}
        \STATE $d=\mbox{distance}(i, j)$
        \STATE $D[i,j]=d$
        \STATE $D[j, i]=d$
  \ENDFOR
  \ENDFOR
 \end{algorithmic}
 \caption{Distance matrix calculation}
 \label{alg:dist_mat}
 \end{algorithm}

The algorithm  takes one sequence $s$ and a window size $w$ as input and generates a distance matrix $D$. The matrix $D$ is initialized with window size $w$ because this is the maximum distance possible between two subsequences. 
Every subsequence has a long overlap with its neighbor subsequences that leads to trivial matches. To avoid trivial matches for each subsequence $s_i$, \cite{yeh2016matrix} suggests excluding a region of length $w$ centered on the starting position of $s_i$.
Since the distance between subsequence $i$ and $j$ ($D[i, j]$) is the same as the distance between subsequence $j$ and $i$ ($D[j, i]$), our distance matrix is symmetric. For time efficiency, we only calculate $D[i,j]$ and assign it to $D[j,i]$.

\begin{figure}
\centerline{\includegraphics[width=0.75\linewidth]{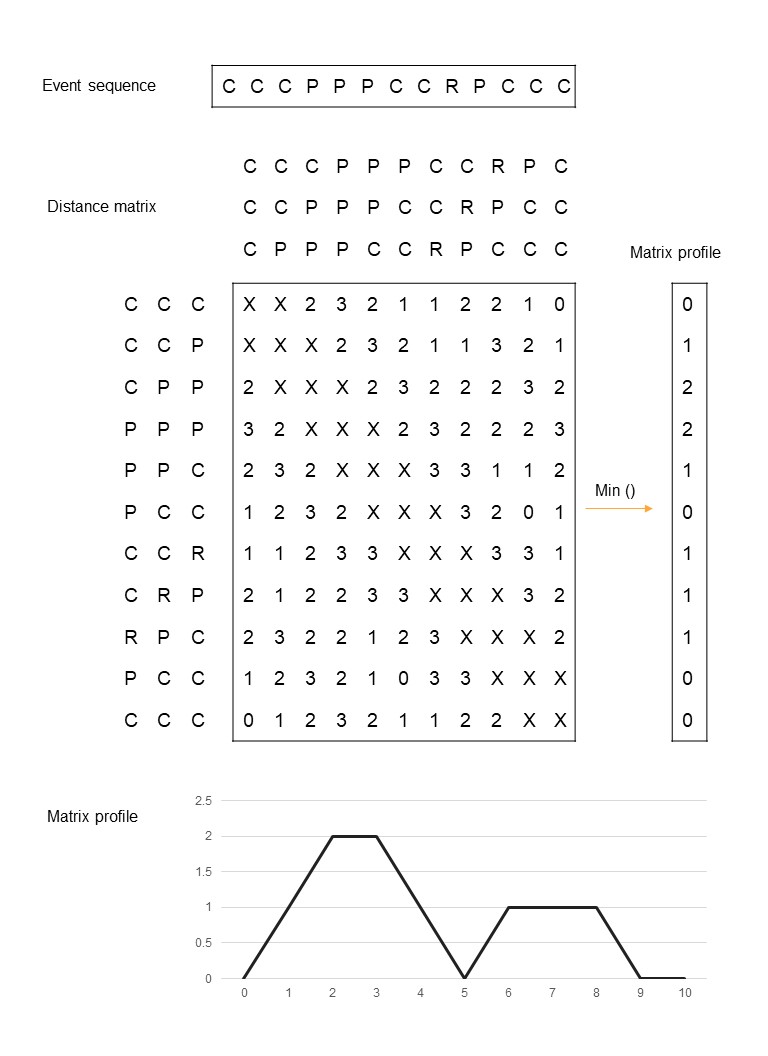}}
\caption{Matrix profile calculation example. From the event sequence, first the distance matrix is calculated. This is a 2-dimensional matrix representing the distance between each pair of subsequences of length $3$ (= window length). The matrix profile is calculated by selecting the minimum value of each row in the matrix which represents the distance to the closest subsequence.} 
\label{fig:mp_example}
\end{figure}

After the calculation of distance matrix $D$, the matrix profile is generated by considering the lowest value of each row as the matrix profile value of that row (Equation \ref{eq:profile}).
\begin{equation} \label{eq:profile}
    P[i]=\min_{\forall j \in n}D[i,j]
\end{equation}
Figure \ref{fig:mp_example} shows an example of the matrix profile calculation for a toy sequence.

Characteristics of matrix profiles such as minimum, maximum, and variance reveal interesting properties of the sequences including motif positions, discord positions, and sequence complexity and novelty \cite{yeh2016matrix}.
To understand what are the differences between groups of sequences, characteristics of matrix profiles of sequences within each group can be aggregated and compared to aggregated characteristics of other groups. For example, if matrix profile average in one group is significantly higher than other groups, it shows that  sequences in that group are less repetitive than sequences in other groups.

\subsection{Sequence Classification}
In this section, we discuss how we converted varied-length sequences to appropriate inputs for existing classification algorithms including SVM and LSTM. 

To use classic machine learning algorithms, we need equal length real-valued vectors.
One way to convert sequences to vectors is using the vectorization methods discussed in Section~\ref{sss:seq2vec}.
We use a k-gram representation for vectorizing the sequences in the dataset before employing a support vector machine (SVM) model to construct a predictive model.

Our second approach for sequence classification is using deep neural network models.
Deep learning models, unlike classic algorithms such as SVM, do not always require direct vectorization. 
These models can have an embedding layer that converts the sequences to real-valued vectors.
We use a Long Short Term Memory (LSTM) model to learn and classify representations for sequences. LSTMs have achieved notable success in natural language processing tasks such as machine translation ~\cite{sutskever2014sequence}. 
Additionally, we test 1-Dimensional Convolutional Neural Networks (1D CNNs) which excel at learning the spatial structure in input data. CNNs are also used in many sequence models such as sentence classification and language translation~\cite{kim2014convolutional}.

We also built a baseline model, which is a simple logistic regression model to be able to compare results of black box models with this glass box baseline. This model considers the length of the input sequence as the only feature for the classification task.

\section{GitHub Sequence Comparison}
This section describes the results of our analysis procedure for studying the  differences between GitHub event sequences.
\subsection{GitHub Dataset}
\label{dataset}
We apply our techniques to the GitHub public events dataset. Our aim is to understand the characteristics of event sequences of human teams versus human-bot teams on GitHub and to  discover if there are differences between these two groups of sequences.
We selected 20,119 software development repositories created in January 2016 that had at least 100 events and more than two human team members. 


We considered a GitHub user a member of a team if they have completed at least one of the following: one push event; five accepted pull requests; ten issue comments; or ten pull request review comments. We have made our dataset available at \cite{supp}.

\subsubsection{Bot Identification}
In our dataset, an account is considered a bot if its type is set to Bot, its name ends with '-bot', and/or it has a high number of identical comments.
In our sample, we identified 304 (1.5\%) teams that had at least one bot.
We denote teams that use automation as \textit{human-bot teams} and teams without bots \textit{human teams}.

\subsubsection{Control for Developers Expertise}
To balance our dataset, we control for developer expertise to make sure the obtained event sequence differences are not due to team expertise.  To do this, we extracted an expertise vector for every team member comprised of 1) number of followers 2) number of following 3) number of public repositories owned by the developer 4) GH-impact score (a measure of influence on GitHub). A developer has a GH-impact score of $n$ if they have $n$ repositories with $n$ stars. This metric is similar to h-index which is used to evaluate impact of scholars.
For each human-bot team, we found the most similar human only team with respect to their expertise vector and downsampled human teams to 304 teams corresponding to the 304 human-bot teams.

\subsubsection{Event Sequence Extraction}
The GitHub activity dataset consists of 14 event types: \textit{push}, \textit{pull request}, \textit{issue comment}, \textit{pull request review comment}, \textit{issue}, \textit{commit comment}, \textit{create}, \textit{delete}, \textit{gollum},  \textit{member}, \textit{public},   \textit{release}, \textit{fork}, and \textit{watch}.
The input vocabulary consists of 14 symbols each corresponding to one event type.
We created the team event sequences using all events, sorted by time, performed within a year of repository creation. 

\subsection{Classifying GitHub Team Type}\label{ss:res_classification}

We trained and tested three different machine learning models for the team type prediction task. 

\subsubsection{SVM}
We used the k-gram representation of sequences and vectorized them using \textit{TF-ISF} model. 
Then, we trained a SVM model using the implementation available in \textit{scikit-learn} machine learning library.

\subsubsection{LSTM}
We used the LSTM recurrent neural network models implemented in \textit{Keras} deep learning library.
Each event was mapped onto a 32 length real-valued vector.
Only the first 1000 events of each team were considered; long sequences were truncated and short sequences were zero padded. As discussed in Section \ref{dataset}, the majority of the teams generate less than 1000 events.
The first layer of our neural network is an embedded layer that uses length 32 vectors to represent each event.
The next layer is an LSTM layer with 100 neurons.
Finally, we added a dense output layer with a single neuron and a sigmoid activation function to make 0 or 1 predictions for the two classes: human team or human-bot team.
Because it is a binary classification problem, we used log loss as the loss function. 
The efficient Adam algorithm is used for optimization. 

\subsubsection{CNN+LSTM}
Our CNN+LSTM model used the same architecture as our LSTM model, but with a 1-dimensional CNN layer and a max pooling layer before the LSTM layer. 

\begin{figure}[h!]
\centerline{\includegraphics[width=0.85\linewidth]{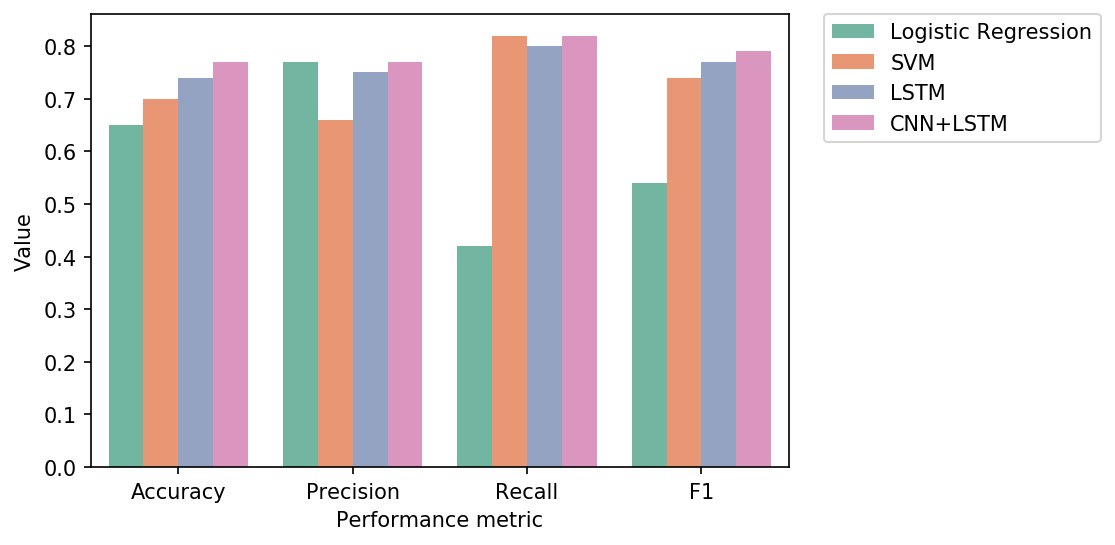}}
\caption{Classification performance of the different models (logistic regression, SVM, LSTM, CNN+LSTM) at recognizing human-bot versus human only team sequences}
\label{fig:seq_classification}
\end{figure}

For the neural network models we hold out 20\% of data for testing and trained the models on rest of the data. 10\% of the training data was used as validation set for tuning parameters.
For the SVM and Logistic Regression models we used 5-fold cross-validation.
Figure \ref{fig:seq_classification} shows the performance of the different classification models. 
Our neural network models achieved the highest F1 scores of 0.79 (precision=0.77, recall=0.82) and 0.77 (precision=0.75, recall=0.80) for CNN+LSTM and LSTM, respectively; while the SVM model achieved F1 score of 0.74 (precision=0.66, recall=0.82). Both models have significantly higher F1 score than our baseline model with F1 score of 0.54 (precision=0.77, recall=0.42), showing that predicting the type of teams is a  non-trivial task and that the sequence of events is helpful in distinguishing team types.
Moreover, these results show that the neural network based black box models perform better than the explainable logistic regression model.

\subsection{GitHub Silhouette Score Analysis}
A k-gram representation of GitHub team event sequences was created using different window sizes $w \in \left \{2, 3, 4, 5 \right \}$. Then \textit{TF-ISF} vectors were extracted for these sequences.

\begin{figure}[h!]
\centerline{\includegraphics[width=0.8\linewidth]{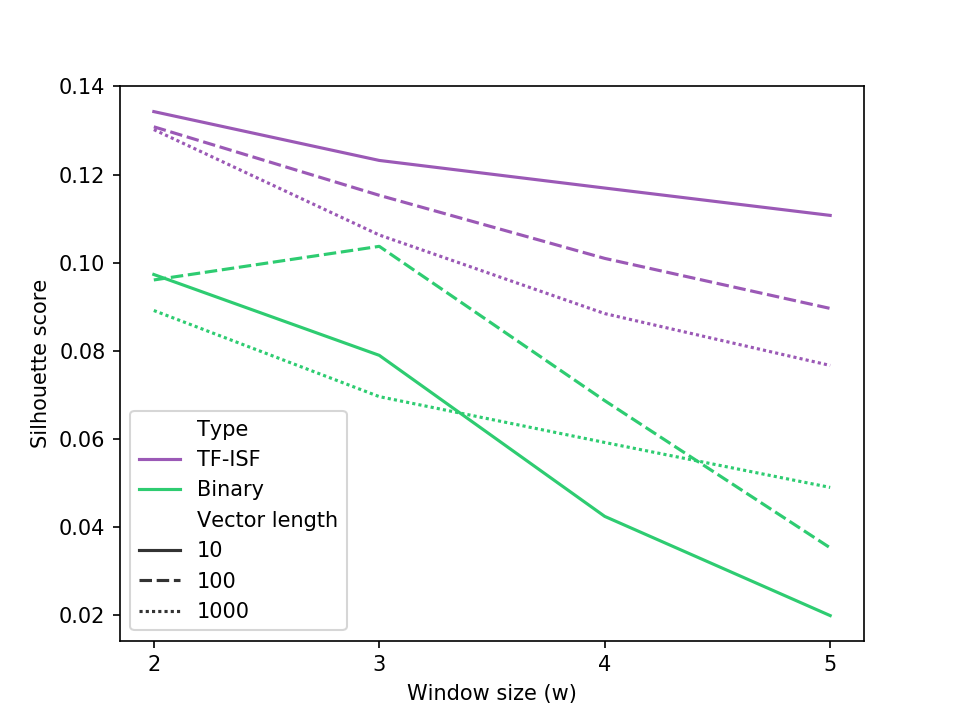}}
\caption{Human vs. human-bot teams silhouette score considering different vector lengths (line style) and different vectorization models (line color).  The best method for detecting differences at all window sizes is TF-ISF with a vector length of 10.  However even the binary vectorization model detects differences between the two groups of sequences.}
\label{fig:var_len}
\end{figure}

To compare the sequences of human teams with the sequences of human-bot teams, we calculated the silhouette score between human-bot team vectors and the downsampled set of vectors of human teams. 
Figure~\ref{fig:var_len} illustrates the amount of distinction between human teams versus human-bot team sequences for different window sizes. Positive values of the silhouette score show that these two groups of sequences are relatively distinct, although they are not completely separate.

The distinction between human and human-bot teams decreases as $w$ increases, where $w$ is the length of the window for constructing subsequences of each sequence.
This occurs because when subsequences become longer, the number of shared subsequences between the sequences decreases.

In our \textit{TF-ISF} model, which creates real-valued vectors from an ordered list of subsequences, the size of the vectors is the  number  of unique subsequences in all sequences.
Since there are many subsequences that are rare, there is an option to limit the size of the vector to only consider most frequent subsequences.
We measured the silhouette score between human and human-bot teams considering vector length to be 10, 100, and 1000. 
Vector length $x$ denotes that a vocabulary of subsequences is constructed that only considers the top $x$ subsequences ordered by term frequency across the sequences.
Figure~\ref{fig:var_len} illustrates the impact of vector length on measuring the distinction. 
The distinction between two groups increases when a shorter vector length is used.
This means that the main difference between human only and human-bot sequences is in the most frequent subsequences.

We also calculated silhouette scores for the binary vectorization model.  This model ignores the frequency of subsequences in order to understand if the distinction between human versus human-bot team sequences occurs because of the difference in frequencies or whether the subsequences themselves also differ.
Green lines in Figure~\ref{fig:var_len} correspond to the binary model. The results show that although using the binary model makes the groups less separated, this model still reveals the distinction between human only and human-bot teams.
This indicates that it is not only the frequency of the subsequences that differs between human only versus human-bot teams but also that different subsequences exist in the these two groups.

\subsection{GitHub Sequences Matrix Profile Analysis}
To understand what is different about human only team event sequences as compared to human-bot teams, we constructed distance matrices and matrix profiles for all sequences.
For the distance calculation between two subsequences, we used Hamming distance due to its time efficiency.

We conducted our experiments for window size of length $20$ based on our empirical analysis.
Table~\ref{tab:mp_hvhb} shows the summary of statistics of matrix profiles of human versus human-bot teams.
The matrix profile consists of distances to the closest subsequence for every subsequence.
The higher average for average of matrix profile values for human-bot teams shows that human-bot teams subsequences are less similar to each other compared to human teams.
That indicates that human teams have more repetitive groups of actions.
A Mann-Whitney U test shows that the average matrix profile value is significantly different in human teams compared to human-bot teams ($p=0.02$). The Mann-Whitney U test was chosen to test the null hypothesis because the data does not follow a normal distribution and a non-parametric statistical test is needed.
Human-bot teams may have less repetitive sequences of human actions if the bots perform repetitive tasks, leaving fewer repetitive series of actions for humans to perform.

Yeh et al. \cite{yeh2016matrix} considered the variance of matrix profile to be representative of the complexity of its underlying time series.
Although human-bot teams have a higher matrix profile variance, the t-test shows that this difference is not significant.
Therefore, we cannot conclude that human-bot teams have more complex sequences.

The minimum and maximum values in the matrix profile are related to motif and discord of the sequence~\cite{yeh2016matrix}. 
There is no significant difference between minimum and maximum values of matrix profiles in human teams versus human-bot teams (\textit{p-value}$=0.1$).

\begin{table}[]
\centering
\begin{tabular}{|l|l|l|l|}
\hline
Metric          & Human-bot & Human & p-value \\ \hline
Variance  & 4.8       & 4.6   & 0.1    \\ \hline
Average & 6.9       & 6.1   & 0.02   \\ \hline
Minimum  & 2.0       & 2.2   & 0.1    \\ \hline
Maximum  & 11.8      & 11.0  & 0.1  \\ \hline
\end{tabular}
\caption{Human versus human-bot teams matrix profile summary. 
}
\label{tab:mp_hvhb}
\end{table}

\begin{figure}[]
\centering
\includegraphics[width=\linewidth]{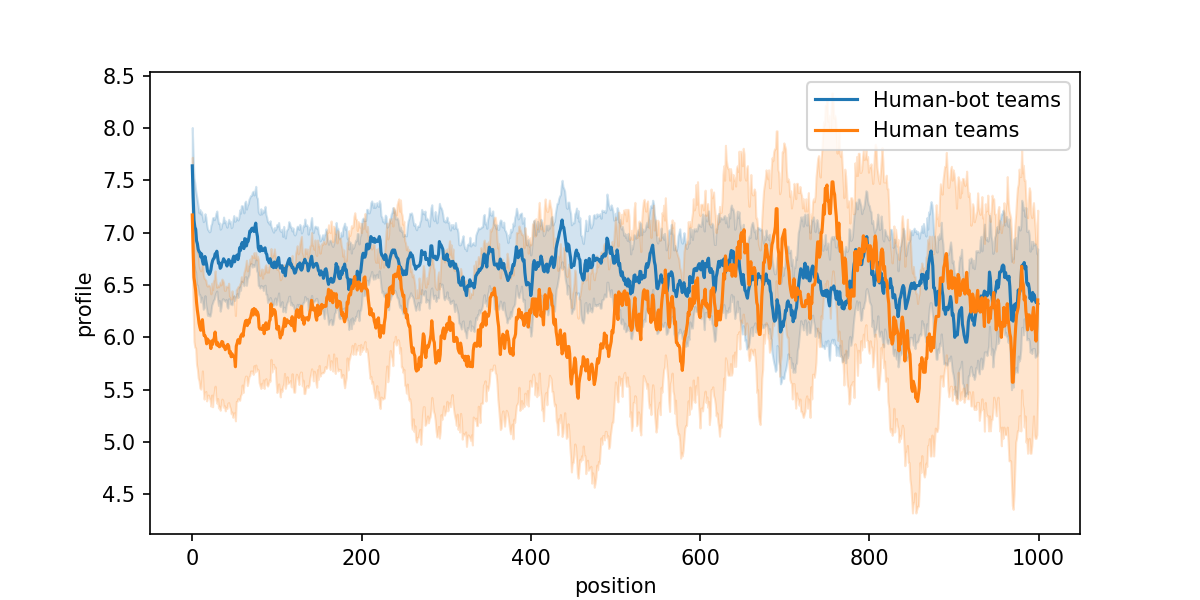}
\caption{Aggregate matrix profile for human-bot teams vs. human only teams.  The human-bot teams have higher values compared to human teams, indicating the higher novelty of sequences in human-bot teams.}
\label{fig:agg_matrix_profiles}
\end{figure}

Figure~\ref{fig:agg_matrix_profiles} shows the profiles of human and human-bot team sequences. We created matrix profiles for full length team sequences but for better visualization, we plotted only 1000 first positions of matrix profiles in Figure \ref{fig:agg_matrix_profiles}. Since 97\% of teams have sequences shorter than 1000, this visualization contains the full length matrix profile of the majority of the teams.
The matrix profile of human-bot teams clearly have higher values compared to human teams, indicating the higher novelty of sequences in human-bot teams.

Figure \ref{fig:agg_matrix_profiles} shows that although human teams have lower matrix profile values at the beginning 
due to simpler,
repetitive groups of actions, as  the human team projects progress, the value and fluctuation of their matrix profiles increases
which indicates that they are becoming more complex. 
However, human-bot teams seem to be complex from the beginning, and they maintain the complexity of their sequences as the projects progress.

\section{Minecraft Action Sequences}
\subsection{Minecraft Dataset}
We used a dataset collected by the Heapcraft project across multiple servers~\cite{muller2015statistical}. The dataset contains two months of data from 45 players, forming 14 person-days worth of active game-play. 
The benefit of this dataset is that it provides ground truth Minecraft actions for collections of raw events.
At random intervals, players were asked to specify the high-level actions they are performing: explore, mine, build, and fight.

Several of the events were excluded by \cite{muller2015statistical}  from the event log due to low frequency, correlation to other events, and redundancy.
Moreover, move, sprint and sneak events were transformed to their corresponding distance or duration. 
We followed the event cleaning procedure presented by \cite{muller2015statistical}  except move, sprint and sneak events were also removed as distance and duration cannot be easily converted to symbols in sequences.

The original study considered the duration of each action to be two minutes centered around the time of response received.
These two-minutes intervals (labeled with high level actions) were used to construct our action sequence dataset.
The sequence dataset of players was created by considering all the events performed by players during the data collection period.
We created the event sequences by assigning a symbol to each Minecraft event and creating an ordered list of symbols for each data point. Since our method relies solely on the order of events rather than their frequencies, consecutive repetitive events are replaced by one event. For example, \textit{aaabbcccd} is transformed to \textit{abcd}.
Finally, we removed sequences that their length is shorter than five and reached 34, 102, 171, and 260 sequences for the fight, explore, mine, and build actions respectively which creates a dataset of size 567.

\subsection{Minecraft Action Classification}
\begin{figure}[h!]
\centerline{\includegraphics[width=0.9\linewidth]{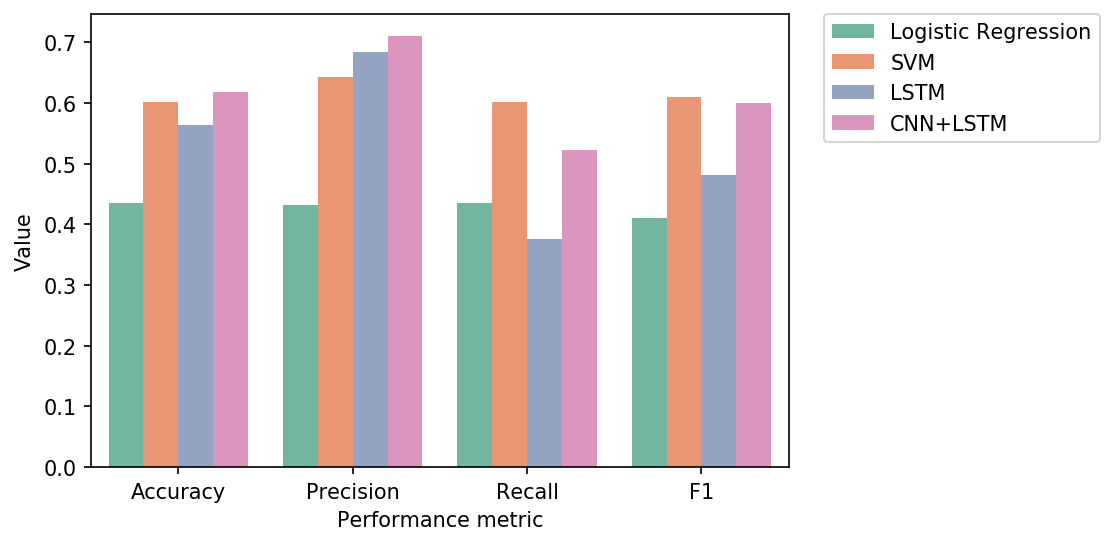}}
\caption{Classification performance of the different models at recognizing Minecraft action sequences.}
\label{fig:minecraft_seq_classification}
\end{figure}

Similar to GitHub sequence classification, we classified Minecraft actions using three different classifiers (i.e. SVM, LSTM, CNN+LSTM) and a baseline (i.e. Logistic Regression based on sequence length).
Figure~\ref{fig:minecraft_seq_classification} demonstrates the performance of the classifiers.
On this dataset, SVM performed better than other classifiers with an F1 score of 0.61 (precision=0.64, recall=0.60). The SVM performance is slightly higher than CNN+LSTM with an F1 score of 0.60 (precision=0.71, recall=0.52).
Note that this is multi-class classification problem with four possible outputs. Therefore, a classifier that randomly assigns label to sequences will achieve an accuracy of 0.25.

\subsection{Minecraft Actions Silhouette Score Analysis}
We vectorized Minecraft action sequences using \textit{TF-ISF} approach as it is superior to the binary vectorization in detecting differences.
Silhouette scores for Minecraft action vectors were calculated using various vector sizes and window lengths.
A window length of $2$ and vector size of $50$ delivered the highest silhouette score of 0.052.
This silhouette score is less than silhouette score of GitHub sequences (0.13).
This indicates that the classification of Minecraft actions is more challenging than identifying GitHub repositories with bots, hence the lower performance is unsurprising.

\subsection{Minecraft Actions Matrix Profile Analysis}
\begin{figure}[]
\centering
\includegraphics[width=\linewidth]{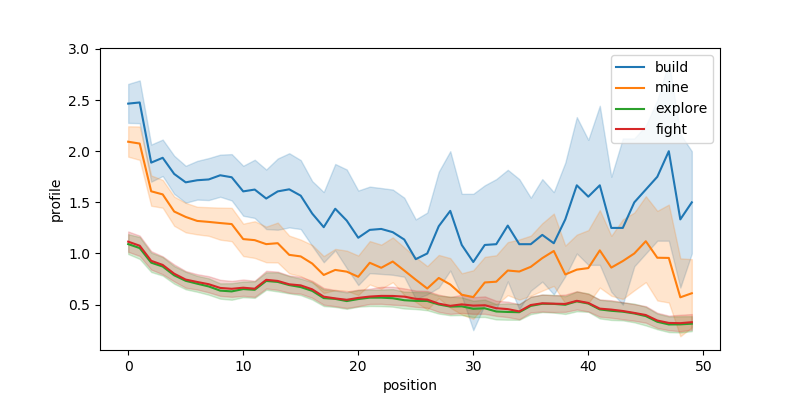}
\caption{Aggregate matrix profile for Minecraft actions}
\label{fig:minecraft_agg_mp}
\end{figure}

We created matrix profiles for all sequences, using a window size of 5.  Figure~\ref{fig:minecraft_agg_mp} illustrates aggregate matrix profiles for each type of action.
Since the matrix profiles of the build and mine actions have larger values at the beginning, we can infer that these two actions start with a subsequence that does not repeat later. 

We calculated minimum, maximum, variance, and mean for matrix profiles of Minecraft action sequences. Table~\ref{tab:minecraft_mp_stats} shows the average of these statistics for different actions.
The variance of matrix profiles is similar across  Minecraft actions. This means that sequences of different actions have similar levels of complexity.
Minimum, maximum, and mean are different in build compared to other three actions, and this difference is statistically significant (\textit{p-value} $<$ 0.05). The same holds true for mine action. 
The higher matrix profile mean in build and mine indicates that these two actions have less repetitive subsequences.
According to Table~\ref{tab:minecraft_mp_stats}, explore and fight have the lowest average minimum, suggesting that these two actions have extremely strong motifs.
The highest average maximum in build action indicates that discords in build sequences are exceedingly distinct from the rest of the sequence.

\begin{table}[]
\centering
\begin{tabular}{|l|l|l|l|l|}
\hline
Metric&Build&Explore&Fight&Mine\\
\hline
Variance&0.5&0.6&0.6&0.5\\
\hline
Mean&2.6&0.9&0.9&2.1\\
\hline
Minimum&1.9&0.3&0.3&1.5\\
\hline
Maximum&3.5&2.6&2.7&3.2\\
\hline
\end{tabular}
\caption{Matrix profile statistics of Minecraft actions}
\label{tab:minecraft_mp_stats}
\end{table}

\section{Conclusion}
This paper proposes an approach to classify groups of discrete sequences and quantify the differences between them.
We present case studies of how our approach can be used to understand GitHub teams and Minecraft actions.
We used our approach to study two different GitHub repository groups: those who use automated accounts (bots) and those who don't.  Our analytic approach reveals subtle differences in teamwork patterns that are difficult to distinguish from event distributions. We believe that sequences of GitHub events can be mapped to team cognitive processes such as knowledge-building, information sharing, and problem-solving; normally in psychology experiments this mapping is accomplished by human observers but we aim to do it with machine learning. 


Our experiments reveal that human team event sequences are relatively distinct from human-bot teams in terms of the  existence and frequency of short subsequences.  This shows that the cadence of activity in human-bot teams is different than human only ones. The matrix profile analysis shows that human-bot teams exhibit differences in both average and absolute maximum values. 
By analyzing the matrix profile of teams, we see that human-bot teams are less likely to repeat event subsequences than human only teams.  Although it is unsurprising that human developers avoid repetition, it is interesting that the usage of bots can be detected from the event sequences alone, without using features from the comments, repository profiles, or code.


Moreover, we utilized our approach to study Minecraft action sequences.
Our analysis shows that build and mine actions have less repetitive subsequences compared to fight and explore.
Sequences of different actions have similar levels of complexity.
Our experiment reveals that improving the performance of Minecraft action classification is challenging because these groups of actions are extremely similar in terms of the existence and frequency of subsequences.

\section{Acknowledgement}
This material is based upon work supported by the Defense Advanced Research Projects Agency (DARPA) under contracts numbers W911NF-20-1-0008 and FA8650-18-C-7823. Any opinions, findings and conclusions or recommendations expressed in this material are those of the authors and do not necessarily reflect the views of DARPA or the University of Central Florida.

\bibliographystyle{IEEEtran} 
\bibliography{main}

\end{document}